\def\USETIMES{false}

\documentclass[letterpaper,twocolumn,10pt]{article}
\usepackage{usenix,epsfig,endnotes}

\usepackage{listings}
\usepackage{color}
\usepackage{graphicx}
\usepackage{ifthen}
\usepackage{subfig}
\usepackage{cite}
\usepackage{url}
\usepackage{balance}

\usepackage[T1]{fontenc}
\usepackage[utf8]{inputenc}

\usepackage{layouts}
\usepackage{multicol}
\usepackage[dvipsnames]{xcolor}
\usepackage{xspace}
\usepackage{paralist}

\ifthenelse{\equal{\USETIMES}{true}}{\usepackage{times}}{}

\usepackage{pifont} %

\makeatletter
\renewcommand\large{\@setfontsize\large\@xiipt{15\p@}}
\makeatother
\usepackage{microtype}

\newcommand{\afblock}[1]{\noindent{\textbf{#1 }}} 

\begin{document}

\newif\ifcomments
\commentsfalse

\ifcomments
\newcommand{\todo}[1]{\textbf{\color{red}{#1}}}
\newcommand{\fs}[1]{\textbf{\color{ForestGreen}{#1$_{flo}$}}}
\newcommand{\hr}[1]{\textbf{\color{WildStrawberry}{#1$_{helge}$}}}
\else
\newcommand{\fs}[1]{}
\newcommand{\hr}[1]{}
\newcommand{\todo}[1]{}
\fi

\definecolor{mygreen}{rgb}{0,0.6,0}
\definecolor{mygray}{rgb}{0.5,0.5,0.5}
\definecolor{mymauve}{rgb}{0.58,0,0.82}

\newcommand{\name}{Santa\xspace}

\newcommand{\HTTPfactor}{2.5\xspace}
\newcommand{\DNSfactor}{5.5\xspace}
\newcommand{\MEMfactor}{2.1\xspace}

\newcommand{\HTTPpercent}{150\,\%\xspace}
\newcommand{\DNSpercent}{450\,\%\xspace}
\newcommand{\MEMpercent}{110\,\%\xspace}

\newcommand{\etal}{et~al.}

\title{Application-Agnostic Offloading of Packet Processing\thanks{This is an extended version of our work presented in~\cite{hohlfeld2018santa}.}}

\author{
\textsuperscript{$\dagger$}Oliver Hohlfeld, \textsuperscript{$\dagger$}Helge Reelfs, \textsuperscript{$\dagger$}Jan R\"uth, \textsuperscript{$\ddagger$}Florian Schmidt,  \\\textsuperscript{$\dagger$}Torsten Zimmermann, \textsuperscript{$\dagger$}Jens Hiller, \textsuperscript{$\dagger$}Klaus Wehrle \\
\vspace{.3em}
\textsuperscript{$\dagger$}COMSYS, RWTH Aachen University \hfill \textsuperscript{$\ddagger$}NEC Laboratories Europe  \\
\textsuperscript{$\dagger$}\{lastname\}@comsys.rwth-aachen.de \hfill \textsuperscript{$\ddagger$}florian.schmidt@neclab.eu\\
}

\maketitle

\subsection*{Abstract}
As network speed increases, servers struggle to serve all requests directed at them.
This challenge is rooted in a partitioned data path where the split between the kernel space networking stack and user space applications induces overheads.
To address this challenge, we propose \name, a new architecture to optimize the data path by enabling server applications to partially offload packet processing to a generic rule processor.
We exemplify \name by showing how it can drastically accelerate kernel-based packet processing---a currently neglected domain.
Our evaluation of a broad class of applications, namely DNS, Memcached, and HTTP, highlights that \name can substantially improve the server performance by a factor of \DNSfactor, \MEMfactor, and \HTTPfactor, respectively. 

\ifthenelse{\equal{\USETIMES}{true}}{}{\vspace{-0.2cm}}

\section{Introduction}

Increasing line rates challenge the packet processing performance of current server systems.
These performance challenges can be attributed to two main overhead factors in network stacks: {\em i)} memory allocations as well as copy operations, and {\em ii)} overheads by performing system calls and the required context switches~\cite{larsen:latency, netmap, sandstorm}.
These costs are particularly significant at high line rates (e.g., multiple 10\,G interfaces), but are already apparent at lower rates if many small requests need to be processed, e.g., for DNS traffic.
Thus, current OS {\em data path} designs significantly challenge packet processing performance in commodity hard- and software where CPU speeds do not scale with increasing line speeds.

The problem of speeding-up server systems is currently addressed by alternative data path designs that entirely bypass the kernel-level network stack as the bottleneck.
One line of research proposes to offload packet processing to dedicated hardware for improved processing performance, e.g.,~\cite{packetshader, arsenic}).
A very active line of research proposes to shift packet processing to user-land stacks and thereby also removes the kernel from the data path, e.g.,~\cite{netmap, Honda14}.
Performance improvements by this approach can be attributed to {\em i)} omitted copy operations and context switches between user space and kernel space and {\em ii)} benefits due to optimized and tailored microstacks. 
Realizations of HTTP and DNS servers as example applications utilizing user-land networking showed drastic performance increases, e.g.,~\cite{Honda14, sandstorm}.

In this paper, we describe a different strategy to accelerate server systems by proposing a widely applicable data path architecture that can benefit from current bypassing approaches, but does not necessarily require the abandonment of well-maintained kernel stacks. 
Inspired by SDN, we propose to split the current data plane into a control and data plane,
by allowing applications to partially offload their packet processing into an application-agnostic rule processor which we refer to as \name.
This rule processor handles frequent requests on behalf of the application and therefore short-cuts the data path.
It can reside in various parts of the network, e.g., in middleboxes, where it can take advantage of current stack bypassing techniques to accelerate packet processing.
However, it can also reside within traditional kernel stacks. From there, it can accelerate packet processing by avoiding costly copy operations and context switches that challenge server performance in the first place.
Based on or examples to accelerate kernel-level packet processing, we show how the optimization of existing stacks---a domain that is currently neglected---can provide competitive alternatives to radical bypassing approaches.
 
The main contributions of this paper are as follows:\\
\begin{itemize}
\vspace{-0.54cm}
	\item We present \name, a widely applicable data path design involving an application-{\em agnostic} architecture allowing user-level applications to offload replies to common requests to a generic rule processor, e.g., to accelerate kernel-level stacks.
	Applications only require minimal changes to benefit from \name.
	\item We highlight the benefits of \name accelerating a broad class of applications by focusing on DNS, Memcached, and HTTP as major Internet applications, covering both UDP and TCP. 
	Our evaluations are based on real world examples, i.e., Facebook's Memcached deployment, ISP-level DNS traffic traces, and static HTTP object properties.
	In all scenarios and without bypassing techniques, \name yields drastic performance improvements.
	Specifically, \name increases the number of processed requests per second by up to a factor of \DNSfactor for DNS, \MEMfactor for Memcached, and \HTTPfactor for HTTP.
\vspace{-0.1cm}

\end{itemize}

By applying \name to the kernel, we unlock the speed of kernel space networking for legacy server software without requiring extensive changes or specialized implementations.
We thus pave the way for new packet processing pipelines and complement the ongoing discussion on user-level network stacks and kernel-bypassing techniques.

\section{\name Architecture}
\label{sec:architecture}

\begin{figure}
	\includegraphics[width=\columnwidth]{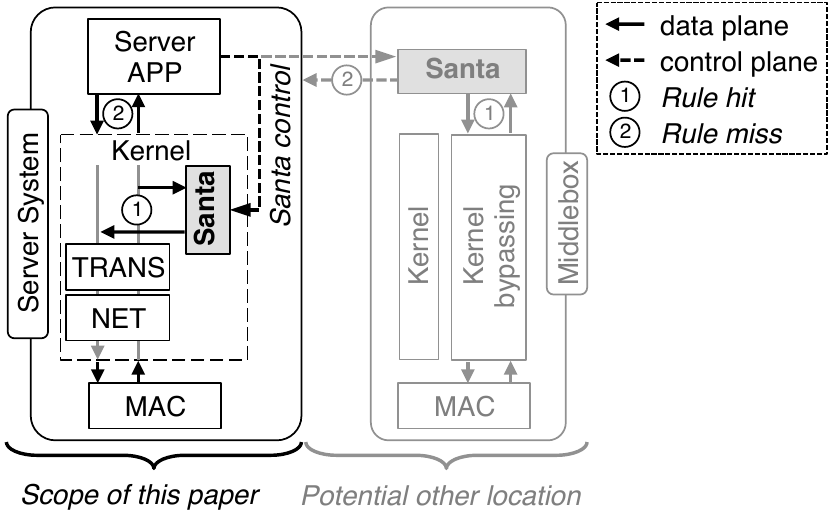}
	\caption{Example execution positions of \name: in-kernel (focus of this paper) and in middleboxes (e.g., accelerated with bypassing techniques such as DPDK or netmap).}
	\label{fig:positions}
	\ifthenelse{\equal{\USETIMES}{true}}{}{\vspace{-.4cm}}
\end{figure}

We accelerate packet processing by splitting it into a \emph{data} and a \emph{control} \emph{plane}, similar to SDN.
This split is based on short-cutting the traditional data plane by inserting an application-agnostic rule processor (\name) that allows applications to partially offload their application processing logic.
\name can reside in various parts of the network, e.g., in the kernel or in middleboxes (cf. Figure~\ref{fig:positions}).
In case of the latter, \name presents a light-weight in-network processing architecture that complements rather heavy-weight NFV-based solutions.
Applying \name to the kernel has the potential to drastically accelerate traditional packet processing by circumventing context switches and copy operations. This domain complements rather radical proposals to bypass the kernel entirely (see Section~\ref{sec:relatedwork}).
By applying \name to the Linux kernel, we show that such approaches are not always necessary, which provides a trade-off decision to application developers and operators.

Unlike caching infrastructures, the control over the offloaded rules remains at the applications using \name.
That is, we provide a control plane offering applications the ability to add, modify, and delete rules on the \name rule processor.
It can also be used for querying the rule processor, e.g., to retrieve access statistics.

\name is most beneficial if a significant part of an application's workload consists of repeatedly serving the same requests with fixed responses (e.g., DNS, HTTP, Memcached, or database workloads).
This highlights the main focus of \name: it will only work well for (temporarily) static request--response pairs, but in those cases, we show that it works exceedingly well.
Furthermore, we present use-cases that show the relevance of such static serving even in today's Internet full of highly dynamic content.

\subsection{Rules: Definition and Expressiveness}
The offloaded rules comprise a \textit{condition} that checks whether an incoming packet matches an anticipated request, and \textit{processors} that construct a reply in that case.
For each packet that is destined for an application that uses \name, the rule processor checks whether the packet matches a rule assigned to its socket.
If a rule matches, \name replies with the offloaded response instead of forwarding the packet to the application.
Otherwise, the packet is handed over to the application which handles the packet. This happens e.g., via the standard socket interface in case of a kernel based implementation, or via a dedicated control channel as in SDN.

\begin{figure}

\includegraphics[width=\columnwidth]{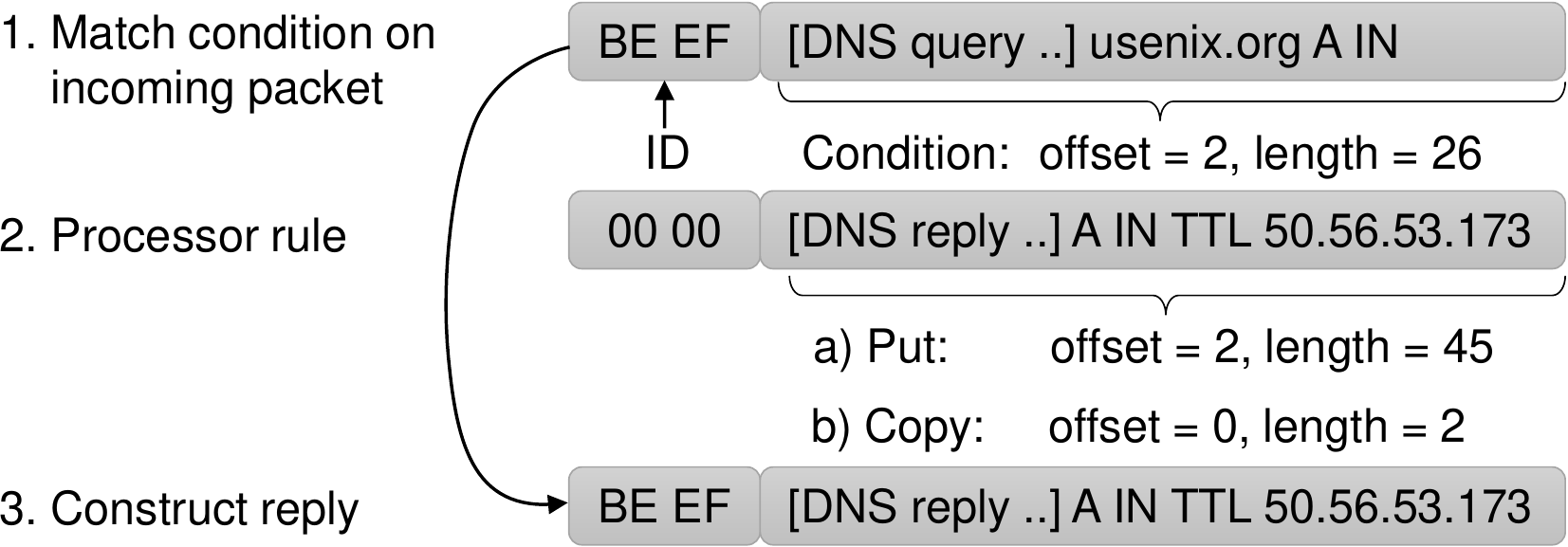}
\caption{Example of a DNS processing rule. In case of an \texttt{A} record request for \texttt{usenix.org}, the rule will construct a response with two successive put and copy processors. The {\em put} operation will put the static part (\texttt{A} record) into the response and the {\em copy} operation will paste the query-specific ID into the response.}
\ifthenelse{\equal{\USETIMES}{true}}{\vspace{-1em}}{\vspace{-.4cm}}
\label{tab:rulesexample}
\end{figure}

Our current condition and processor implementations includes an offset, a length, and the pattern to be matched or a simple data manipulation instruction.
For example (see Figure~\ref{tab:rulesexample}), a DNS server would create a condition that recognizes packets querying a certain \texttt{A} record for which the response is constructed with successive processors: one that contains the reply (e.g., the IP address of the domain), and one that copies the transaction ID identifying the request into the response.

This very basic functionality has several advantages.
By only allowing very simplistic and specific operations when matching requests and constructing replies, we circumvent security issues that are bound to appear if we put complex tasks into critical locations (e.g., kernel space).
Furthermore, the simplicity of these building blocks allows us to keep the rules application-\textit{agnostic} and more importantly results in low computational overhead.

We have currently implemented two processor types: 1) \emph{copy}--copy from the incoming packet and 2) \emph{put}--copy from a predefined template.
Furthermore we implemented one condition type, a hash-based content comparison.
In our evaluation, we show that this simple ruleset suffices to offload responses for such diverse applications as DNS, HTTP, or Memcached.

\subsection{Efficient Rule Matching}
\label{sec:hashing}

As \name targets high-performance applications, handling high packet rates is a fundamental requirement.
Consequently, we must use an efficient solution for determining whether or not a condition matches.
We decided on using a hashing-based approach as it is commonly used for pattern matching.

\begin{figure}
\centering
\includegraphics[width=\columnwidth]{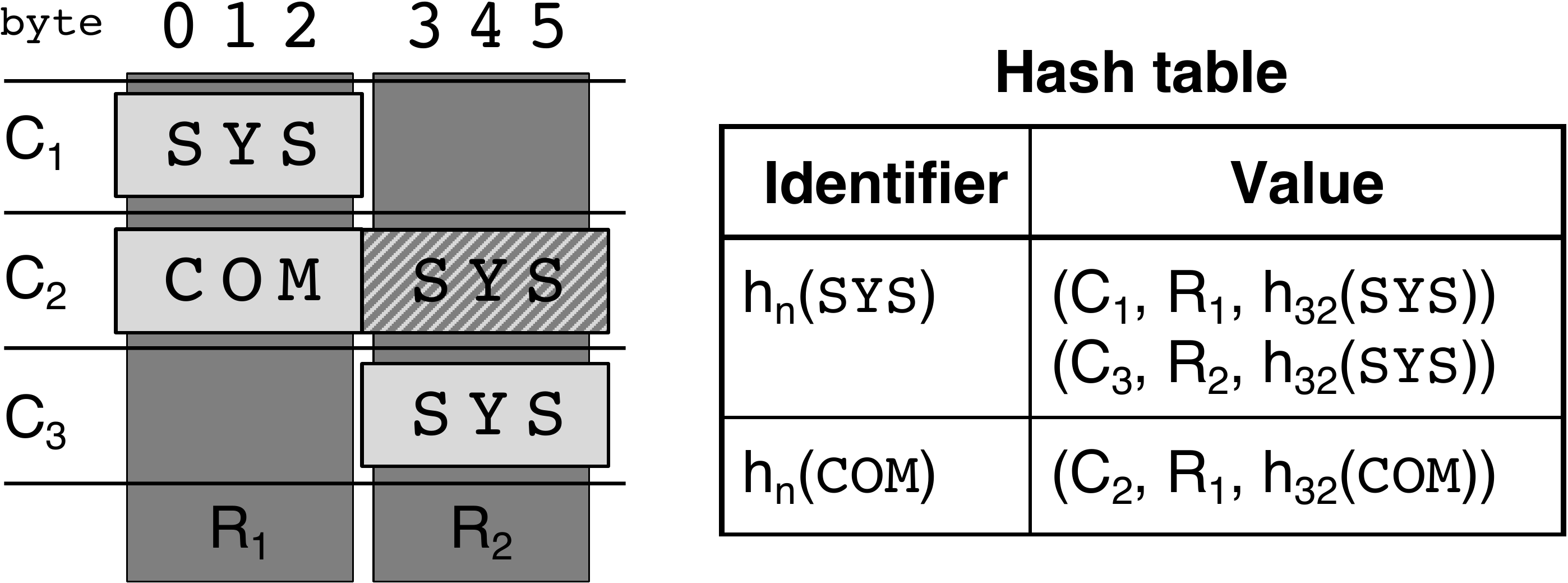}
\caption{Conditions $C_1,C_2,C_3$ are inserted into \name. The algorithm identifies two regions ($R_1, R_2$) for matching. For each condition's starting region, the 32-bit FNV-1a hash ($h_{32}$) is calculated and put into the hash table. Here, we keep track of the condition, region, and $h_{32}$ value for match verification.}
\ifthenelse{\equal{\USETIMES}{true}}{}{\vspace{-.3cm}}
\label{fig:hashing}
\end{figure}

However, \name should handle conditions with arbitrary offsets and lengths.
Thus, a naive hashing-based approach, i.e., hashing each condition's pattern, would require us to calculate many hashes for each incoming packet (one hash for each offset-length combination that occurs in at least one condition).
To reduce this overhead, \name identifies small, non-overlapping \textit{regions}, each defined by an offset and a length, and calculates hashes only for these regions.
Then, for each rule, \name assigns the first region that shares the offset with the rule and that is smaller or equal in length compared to the whole condition to be matched.
For this, we calculate the hash of the (sub-)pattern (induced by the region on the pattern to be matched) and store a reference to the full condition and to the corresponding region in a hash table (cf. Figure~\ref{fig:hashing}) using the calculated hash as the identifier.
This allows us to identify conditions that could match to reduce the number of required full condition checks.

We explain the algorithm by means of an example given in Figure~\ref{fig:hashing}.
In this example, two conditions start at the same offset, and a third starts three bytes later.
Conditions $C_1$ and $C_2$ would yield one shared region $R_1$ with the same offset and the shorter length of both conditions.
$C_3$, however, is responsible for producing another region $R_2$.
The conditions $C_1$ and $C_2$ are linked to their starting region $R_1$, whereas $C_3$ is linked to $R_2$.
We store these links in the hash table with the hash value of the condition's region-defined sub-pattern as the identifier.

To check a packet, \name calculates, for each region, the hash of the corresponding packet's sub-part and checks for occurrence of this hash in the hash table.
If such an entry exists, \name performs a byte-wise comparison between the packet and the full condition, as a region may cover only a subset of the full condition and hash collisions may occur.

The used hash table size has a major impact on the amount of collisions.
For an $n$-bit hash, the table size is $2^np$ with $p$ denoting the size of a pointer.
In comparison to a 32-bit hash table requiring 32\,GB, a 22-bit hash table only requires 32\,\emph{MB}.

For our implementation, we use the extremely fast FNV-1a~\cite{fnv} hash.
To match the hash table key length, we cut the hash to the needed key size.

Still, to cope with a possibly large number of hash collisions, we employ a two-stage collision resolution.
First, we calculate a 32-bit hash and keep it for each condition's region-induced sub-pattern.
With this knowledge, we ensure a matching 32-bit hash value.
Second, we also ensure a matching region for each candidate.
For example, the collision in Figure~\ref{fig:hashing} (\texttt{SYS}) cannot be resolved by the 32-bit hash, because the collision stems from two conditions having the exact same hashing input.
However, the collision can be resolved by looking at the regions.
If the hash of \texttt{SYS} was calculated from $R_1$ of the incoming packet, $(C_3, R_2)$ cannot be a valid result.
Note that the same sub-pattern from $C_2$ is disregarded in $R_2$ as the condition does not start here (cf. hatched area Figure~\ref{fig:hashing}).
Finally, the whole condition found in the hash table has to be checked against the incoming packet to ensure that it is a correct match.
This is not only due to potential hash collisions, but also because we compute the hash possibly only on a \textit{subset} of a condition.
For example, an incoming packet that contains the value \texttt{COMS\emph{I}S} in bytes 0--5 would show up as a candidate for $C_2$ by containing the value \texttt{COM} in $R_1$ and is checked.
Nevertheless, it is obviously not a correct match.

Empirical evaluation of our Memchached use-case (cf. Section~\ref{eval:memcached}) shows that the hash table size has an influence on the performance---we can trade memory for performance.
By serving 1\,M keys we observe a performance increase of factor 1.48 between using a 16-bit and a 22-bit hash table.

Still, our hashing system reduces the matching overhead significantly.
In the evaluation, we show that it allows the use of many thousands of rules without significant performance drops.

\section{Unused Potential of Kernel Stacks}
While \name can offload processing to middleboxes (cf. Figure~\ref{fig:positions}), we lay the focus of this paper on \name's in-kernel application in servers.
Server applications largely rely on kernel-level stacks for packet processing.
However, the packet processing performance of current network stacks suffers from a partitioned data path.
In this partitioning, switching from the hardware to the kernel space and from the kernel space to the user space involves costly context switches and copy operations.
These operations can considerably degrade the packet processing performance of server systems, especially for small packets~\cite{larsen:latency, netmap}.
We complement the works that propose to bypass the kernel entirely---and thus remove the kernel from the data path (see Section~\ref{sec:relatedwork})---by short-cutting the data path with \name to drastically accelerate packet processing in the traditional Linux kernel.
Thereby, we highlight that kernel-level packet processing can be accelerated without having to abandon well-maintained and feature-rich stacks from the data path.
We therefore start by motivating why classical packet processing via the kernel is slow, before we describe the kernel-level implementation of \name in Section~\ref{sec:implementation}.

\subsection{The Price of Partitioning}

From a system perspective, the receive data path processing can be split into three separately handled parts:
(1)~At the bottom, the PHY and (often) MAC processing is efficiently done in hardware.
(2)~The central portion of the processing, the network and transport layer, is done in the operating system's network stack.
(3)~Finally, the application is in charge of its own protocols.

This partitioning has the advantage of providing well-defined interfaces between each of the building blocks, e.g., the quasi-standard Berkeley socket interface that has been in use since the 1980s.
However, one large disadvantage is a loss of efficiency.
Between each of the parts, a context switch occurs (an interrupt and a system call, respectively), and data is copied from one memory area to another.
Especially the mode switch between user and kernel space is a large bottleneck.
Depending on the transmitted data, system calls can easily account for a third of the packet processing time within a host\cite{larsen:latency}.

\subsection{Reducing the Partitioning Overhead}
\label{sec:partition}
Reducing the sizable partitioning overhead in packet processing is a core motivation for many works in the field which we discuss further in Section~\ref{sec:relatedwork}.
Approaches that have seen much exposure in recent years are netmap~\cite{netmap} and DPDK~\cite{dpdk}, which form a part of high-performance network server solutions such as Sandstorm\cite{sandstorm} or optimized user space stacks such as Seastar~\cite{seastar}.

The idea of these approaches is to completely bypass the kernel's network stack and instead directly map data from the hardware queues into the application.
The performance gain stems from two sources:
First, data can be directly transferred to and from user space, eliminating one copy operation.
Second, the application is required to not only process application-layer protocols, but also the network and transport layer ones.
Specialized microstacks only containing the required functionality of the specific setup can be further optimized.
They have been shown to support line rate throughput for a wide range of packet sizes and use-cases (see e.g.,~\cite{sandstorm, seastar, jeong2014mtcp}).

However, one of the advantages of bypassing is also a disadvantage.
By requiring the application to provide and use its own network stack for network and transport layer processing, bypassing usually abandons the exceedingly well-tested, well-maintained, and feature-rich network stack available inside the OS.
In fact, the Linux implementation of network and transport layer are very efficient.
This is shown by StackMap~\cite{yasukata16stackmap} which dedicates a NIC to an application for bypassing kernel data reception while still using Linux networking functionality.

Nevertheless, dedication of NICs, new APIs, and different programming paradigms hinder a widespread adoption in well-established networking applications.

\section{\name in the Linux Kernel}
\label{sec:implementation}

Given the potential performance gains, while maintaining compatibility to legacy software, we decided to implement \name for the Linux Kernel 3.18. %
\name comprises a main part, which is independent of the main kernel files and resides in its own subtree, and several hooks in relevant places inside the network stack (i.e., the socket, TCP, and UDP layer).
The footprint of these hooks is quite small, only about 300 lines of code, of which a large portion is due to additional socket options enabling the control plane.
The \name source code will be available to the research community~\cite{santa-source}.

\subsection{Inserting \name into the Receive Path}

Processing of received packets is done in the Linux kernel within the \texttt{NET\_RX} softirq.
Under normal conditions (and also for \name, if no rule matches), the packet is eventually passed into the socket buffer, ready to be read by an application's system call.
If on the other hand, \name finds that the packet matches a rule, it creates a response from the respective processor(s) and then sends out the resulting packet(s).
We detail this process for TCP and UDP later in this section.

For both cases, however, we do not leave the \texttt{NET\_RX} softirq context until the hardware prepared the packet for sending.
Thus, we inevitably increase the time the system spends in softirq context, because the send path is now also traversed during the softirq.
This is not a problem in and of itself; it is merely a result of skipping system calls and thus a key contributor to our performance increase.
However, it is important to understand that this moves the potential bottleneck of the system.
Normally, under high load, the system is not able to reach line rate because it spends a vast amount of its CPU time switching back and forth between application and kernel, copying data between them, and processing requests in user space.
Once we remove this bottleneck, the softirq processing is potentially the next bottleneck to cope with.
This may happen earlier than one might expect because, by default, the \texttt{NET\_RX} softirq is only processed on the CPU the hardware interrupt was triggered on, which, depending on the hardware, might only be a single core in the system.

Thus, we require parallel processing of \texttt{NET\_RX} softirqs to unlock the full potential of \name.
With receive-side scaling (RSS) and receive packet steering (RPS), the kernel already provides such a mechanism.
Both mechanisms allow us to process packets of the same flow on the same CPU, thus eliminating limitations of simple hardware IRQ distribution.
We do not go into further details of RSS and RPS here; for this paper, it is merely important to understand that both allow us to distribute the processing of incoming packets efficiently over several cores, unlocking the full potential of \name.

In the following, we explain in more detail how and where \name matches incoming packets and constructs responses.
The behavior differs between UDP and TCP. We start with the explanation of UDP because it is the simpler case.

\subsection{UDP Processing}

For UDP, the process is straightforward.
After a packet has been associated with an actual socket, we intercept it to check whether the incoming datagram matches a rule.
This is important as applications insert rules for specific sockets, hence, we need to know which socket a packet should be delivered to.
Otherwise, an application could hijack other applications' packets by inserting rules that match those packets.
If a rule matches, \name constructs a reply from the rule's processors, inserts it into the network stack's transmit path and discards the packet that triggered the rule.
That way, it is neither handed over to the application, resulting in a duplicate answer nor triggers the costly traversal of the user--kernel barrier. 

\subsection{TCP Processing}
\label{sec:implementation:tcp_processing}

The TCP case is significantly more involved.
First of all, there is a fundamental difference between UDP datagrams and TCP segments.
While UDP sends out datagrams as they were handed over by the application, TCP data is in the form of a stream and is packetized at the network stack's discretion.
In theory, we therefore would have to aggregate incoming TCP data from packets, and do our matching on the reconstructed stream.
This, however, poses a fundamental question: Where is the start of our match?
Doing a search over the stream for our match is theoretically possible, but practically infeasible due to the large overhead.

\afblock{Matching assumptions.}
To focus on our use-case of offloading the handling of common request--response pairs, we make two important assumptions:
\begin{inparaenum}[\em 1)]
\item We assume that requests fit into a single TCP segment and are not distributed over multiple segments alike TCP Fast Open~\cite{Radhakrishnan:2011:TFO}.
	\item We further assume that one segment only contains a single request and multiple requests are not aggregated by the sender into a single TCP segment.
\end{inparaenum}
These assumptions allow improving the matching performance and keep the matching strategy simple and analog to handling UDP datagrams.

\afblock{Empirical motivation.}
We next empirically motivate our assumptions using both an ISP-level trace and an active measurement by focusing on HTTP as the most prevalent TCP-based protocol~\cite{IXPAnatomy, GregorDSL}. 
The ISP-level trace comprises anonymized HTTP connection logs captured in the residential access network of a major European ISP over the course of 6 days in 2015.
The complementing active measurement comprises header information captured by crawling the Alexa Top 1\,M list on January 1st, 2016 (available at the HTTP archive~\cite{httpArchive}).

Concerning assumption 1) (matching restricted to single segments), we observed that a significant fraction of the monitored HTTP requests fit into a single TCP segment.
Concretely, 86.2\,\% of the HTTP requests in the ISP trace (98.8\,\% for the Alexa crawl) fit into a single packet with MSS 1436\,B (considering typical option sizes).
51\,\% (65.6\,\%) even fit into the default MSS of 536\,B.
Thus, a major share of the observed HTTP requests does not require fragmentation and can be directly matched with \name.
We remark that matching over multiple segments would be possible after byte stream reassembly, but is currently not implemented to optimize the matching performance.
Instead, the current implementation will process large requests in the user space application.

Concerning assumption 2) (at most one request per segment), we observed that most HTTP connections only carry a single request and pipelining is not prevalent, e.g., as modern browsers parallelize and split requests over multiple connections.
Concretely, 74\,\% of all HTTP connections in the ISP trace only involve a single request ($92\,\% \le 4$ requests).
These findings closely resemble traffic pattern observed at the Yahoo! CDN~\cite{YahooCDN}.
In the rare case that HTTP pipelining is used, i.e., multiple requests are batched after each other, we may match one or more requests depending on the packetization to different segments.
However, typical browsers only send an additional request after having received the HTTP response header for the prior request to prevent head-of-line blocking.	
Thus, Assumption 2 reflects typical HTTP traffic.

\afblock{Implementation challenges.}
In contrast to UDP, TCP's complexity poses a set of implementation challenges.
\begin{inparaenum}[\em i)]
	\item TCP's bookkeeping needs to be performed in case of a match, e.g., sequence numbers need to be increased.

	\item Since we optimize for the common case, we hooked into the fast-path and therefore do not match on out-of-order and out-of-window segments processed in the slow-path.
	We remark, however, that matching these corner cases is possible at higher computational costs.
\end{inparaenum}

\subsection{Userspace API}

To allow applications to offload packet processing tasks to \name, we realized a simple user space API.
Once a socket is created, the application can attach a \emph{processor} and a \emph{condition} (cf. Section \ref{sec:architecture}) that define the matching and the resulting action.
We illustrate the usage by an example shown in Listing \ref{fig:santa_usage}.

\lstset{
	language=C, 
	breaklines        = true,
 	breakatwhitespace = true,
	basicstyle=\footnotesize\ttfamily,
	xleftmargin=10pt,
	framexleftmargin=10pt,
	commentstyle=\color{mygreen},    %
	keywordstyle=\color{blue},       %
	stringstyle=\color{mymauve},
	numbers=left,
	numberstyle=\tiny,
	frame=tb,
	columns=fullflexible,
	showstringspaces=false,
	breaklines=true
}
\begin{lstlisting}[label=fig:santa_usage, captionpos=b, escapechar=|, caption=\name usage example -- we match \texttt{Hello} while responding with \texttt{COMSYS} on socket \texttt{fd}.]
#include "santa-userspace.h" /*userspace part*/
struct santa_processor p = {0};
struct santa_condition c = {0};
p.type = SANTA_COPY_TEMPLATE; |\label{line:type}|
p.buf = "COMSYS"; p.len = 6; p.offset = 0; |\label{line:p}|
c.buf = "Hello"; c.len = 5; c.offset = 0; |\label{line:c}|
/*set processor for condition to specific socket*/
c.p = setsockopt(fd, SOL_SOCKET, SO_SANTA_P, &p, sizeof(p));  |\label{line:p_c}|
/*add the condition*/
setsockopt(fd, SOL_SOCKET, SO_SANTA_C, &c, sizeof(c)); 
\end{lstlisting}

After defining the type of the processor, i.e., copy the data from a predefined template (cf. Lines \ref{line:type} and \ref{line:p}), this processor is attached to the socket {\tt fd} via the {\tt setsockopt} system call.
This call returns an identifier of the newly added processor, which can be bound to a condition (cf. Lines \ref{line:c} and \ref{line:p_c}).
Finally, this condition is also attached to the socket; from now on \name will intercept packets matching the condition and replies with the defined answer. 
All other packets destined for this socket not matching the condition are forwarded to the application as usual.
Update and delete methods are handled likewise.

Note that a non-root application can only alter \name properties of its own sockets.

\section{Evaluation: DNS Server}

\name allows {\em frequently accessed} and (temporarily) static content to be served at {\em lower latency} and {\em higher throughput} in the number of requests.
We demonstrate this ability for three widely-used applications for the two relevant transport protocols, i.e., UDP and TCP. 
In this section, we start with using \name to accelerate UDP packet processing, since this is the simpler case.
As an example application, we focus on accelerating DNS as widely-used UDP-based application~\cite{ixpappmix}.
Since performing name resolutions are the first steps in many Internet transactions, optimizing DNS performance helps to optimize the performance of Internet-based applications.
In particular, drastic increases in throughput allow highly loaded servers to reduce the overall load and to serve a much larger number of clients with the same hardware.

\subsection{Testbed Setup}
\label{sec:dns:testbed}
We evaluate the performance of \name in a testbed study.
Our evaluation testbed consists of a single server, Quad-Core Intel i7 CPU at 3.6\,GHz, 16\,GB of RAM.
This server runs the \name Linux kernel and a BIND 9.10.2 DNS server which we extended to utilize \name. 

Four load generating clients are connected via 10\,G Ethernet over a Netgear switch to the server.
The selected number of clients allows creating an overload scenario by fully utilizing the links.
The reason to focus on a high-load / overload scenario is that this challenges the performance of traditional user space packet processing the most.
Due to the small packet size of DNS requests, we introduce a high amount of per-packet processing overhead.
Therefore, we expect performance optimizations to be the most pronounced in this region.

Our load generation is based on replaying DNS requests according to pre-configured popularity distributions using \emph{DNSPerf}\cite{dnsperf}.

\subsection{Baseline Performance: BIND}
We start by showing that our \name extension has no performance drawback over an unmodified vanilla Linux kernel in the absence of matching rules.
That is, we compare the baseline performance of our modified kernel to the unmodified kernel when no \name rules are installed and all requests are handled by BIND. 
First, this evaluation provides an intuition on the achievable performance in our testbed with unmodified standard software.
Second, it shows whether our kernel modification has negative performance implications on the standard kernel.

To this end, we measure the BIND 9 performance on a vanilla Linux kernel and our modified \name kernel striving for a maximum number of replies.
We configured BIND to serve only a simple DNS request for a single \texttt{A} record.
We expect that this minimum setup results in the least processing overhead for BIND, providing us with an upper performance bound.
For the setup using the modified \name kernel, no rule is configured, so all requests are passed through to BIND.
Thus, no \name acceleration is in place, which ensures that we only compare the performance of the unmodified to the modified kernel.

As workload, we generate DNS requests to the resource record (RR) pre-configured in BIND.
We then measure the performance over 45\,s intervals using a warmup period of $\approx 15$\,s to measure in a more stable region.
As a performance metric, we measure the served requests per second. 
We repeat the experiment 30 times for both kernels.

Our results show that both kernels provide similar performance (c.f. Figure~\ref{fig:eval:alexalinear}, 1 zone entry bars).
We observe that the maximum performance of BIND in our setup is $<$ 640\,k replies per second is equal for both kernels showing that the \name extension has no negative performance implication on the standard Linux packet processing performance.

\subsection{Baseline Performance: \name}
\label{sect:santaDnsBaselinePerf}

\begin{figure}
	\centering
	\includegraphics[width=\columnwidth]{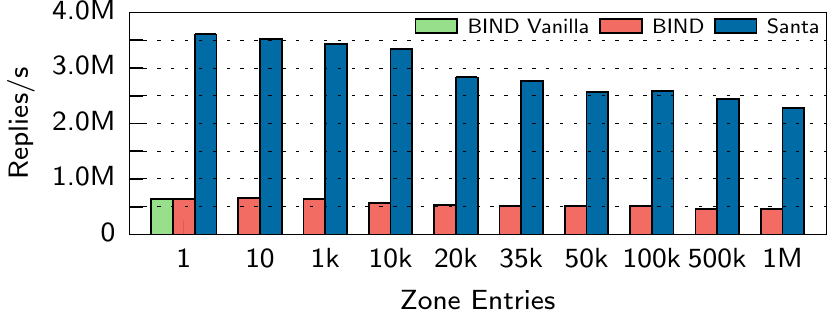}
	\caption{Compared to BIND, serving all requests with \name increases throughput by factor 5 throughout any number of installed rules. The throughput however decreases with higher amounts of rules.
	}
	\label{fig:eval:alexalinear}
	\ifthenelse{\equal{\USETIMES}{true}}{}{\vspace{-10pt}}
\end{figure}

We next move to evaluate the achievable performance of our \name extension.
For this, we configure the server to respond to all incoming DNS requests with \name.
To use a realistic request structure (i.e., length and domain pattern), we use domain names from the \textit{Alexa} top 1\,M list. %
This request pattern serves as an example workload of a DNS resolver, where each entry is equally popular (we show a more realistic, power-law distributed, scenario in Section~\ref{sec:rwdnstraffic}), giving insights about the performance impact of the number of rules installed into \name.
As a request pattern, each client requests a permutation of $n$ unique DNS names using DNSPerf to saturate the server.
On the server side, we install rules for these $n$ DNS names in \name.
Here, we use a 22\,bit hash table to mitigate hash collisions.
Again, we measure the number of served requests per second.

Figure~\ref{fig:eval:alexalinear} shows the sum of served requests per second while increasing the number of rules installed in \name from 1 to 1\,M entries of the Alexa top 1\,M list.
As in the previous section, we measure the performance over intervals of 45\,s while repeating the experiment 30 times for rule set size $n$.\fs{Error bars? Mention that results are so stable that error bars are invisible?}

This evaluation shows a performance increase by a factor of 4.9 to 5.6 as compared to BIND for all tests.
Specifically, up to 10\,k configured hosts, we observe a stable number of 3.3\,M to 3.6\,M replies per second.
The performance starts dropping at 10\,k configured zone entries.
We attribute the observed performance drop to emerging hash collisions.
As all DNS request conditions start at the same offset and a rather short region, domain names with the same prefix result in the same hash value.
Nevertheless, we maintain a generally high performance that clearly outperforms BIND.

This baseline evaluation already shows the potential of our approach.
We remark that this use-case is artificial as no single DNS server will likely serve as many entries and more importantly, not all entries will be equally popular.
Therefore, we will now focus on evaluating \name based on the request pattern of a real-world DNS resolver.

\subsection{Properties of Real-World DNS Traffic}
\label{sec:dnsmeasurements}
To base the evaluation of \name on realistic real-world DNS traffic, we analyze a network trace of end-user DNS request traffic.
The data was captured in a small segment of the residential broadband access network of an ISP over the course of 60\,h in May 2015.

\begin{figure}[t]
\vspace{-13pt}
\subfloat[DNS RR Popularity]{\label{fig:dnspop}\includegraphics[width=0.48\columnwidth]{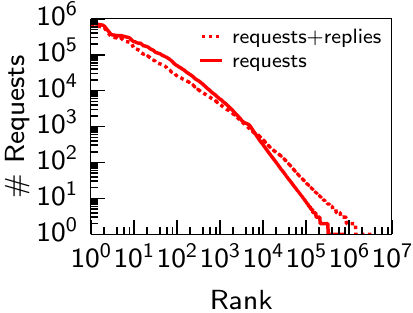}}
\hfil
\subfloat[\name Hit Rate]{\label{fig:cachehit}\includegraphics[width=0.48\columnwidth]{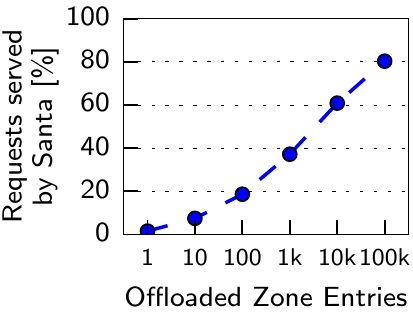}}
\caption{Requests to an ISP DNS server follow a power-law \protect\subref{fig:dnspop}. Thus, handling few heavy-hitters with \name yields high hit-rates \protect\subref{fig:cachehit}.}
\label{fig:dnstrace}
\ifthenelse{\equal{\USETIMES}{true}}{}{\vspace{-1em}}
\end{figure}

In our data, we observe 42.6\,M requests to 727\,k distinct RRs.
The request frequency of the resolved RRs follows a power-law, i.e., very few DNS records receive the bulk of the requests.
Processing heavy-hitters with \name potentially improves the overall performance of a DNS server.
We show the request frequency of each RR ordered by decreasing frequency as solid line in Figure~\ref{fig:dnspop}.

However, as DNS request-to-response mappings are only stable for limited time spans (e.g., minutes in the case of CDNs), changing mappings require updates to the corresponding \name rules.
To understand how many requests can be satisfied with one mapping before an update is necessary, we investigate combinations of resource record, record type, and answer.
This grouping results in 3.6\,M distinct combinations also following a power law, depicted by the dashed line in Figure~\ref{fig:dnspop}.

The observed power-law suggests that stable mappings receive substantial hits before rule updates are required.
Thus, the overall DNS server performance can be significantly optimized by an accelerated processing of heavy-hitters.
By assuming an optimal offloading strategy of a-priori populating \name with the $n$ most popular objects, we see that offloading only 100 of these DNS RRs already enables \name to accelerate 18.6\,\% of requests (cf.\ Figure~\ref{fig:cachehit}).
Increasing the amount of \name rules to include the top 10\,k requested DNS records already yields a hit rate of 60.8\,\%.
We show \name's benefit for this use-case by a testbed-driven evaluation in the next section.
Note that authoritative and root DNS servers offer an even greater potential to benefit from \name acceleration due to lower and more stable set of served RRs\hr{<-- koennte noch imho raus}.

\subsection{Applying \name to Real DNS Traffic}
\label{sec:rwdnstraffic}
We now evaluate \name with a realistic workload pattern derived from the trace (cf. Section \ref{sec:dnsmeasurements}).
This workload pattern allows generating different mixtures of traffic served by BIND {\em and } \name.
That is, we configure BIND to offload the $n$ most popular records to \name.

We base this evaluation on the same testbed setup as used in the previous evaluations.
However, this time, regardless of the offloaded request set, each client picks requests from the Alexa 1\,M hosts at random, weighted according to the probability distribution observed in the previously presented ISP trace (see Figure~\ref{fig:dnspop}).
As a result and unlike our previous evaluations, only the offloaded entries are served by \name, whereas the remaining traffic is served by BIND.
Due to DNS trace anonymization, we establish a canonical mapping between the hashed DNS names and hosts in the Alexa top 1\,M list. %
Thus, each client issues requests from the entire domain set; however, the frequency of each requested host follows the measured power-law distribution.

\begin{figure}
	\includegraphics[width=\columnwidth]{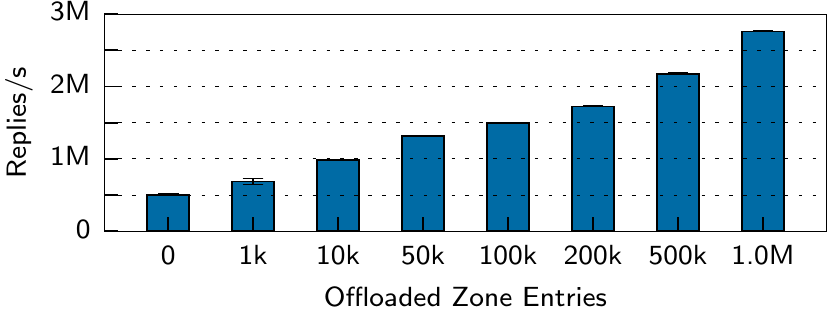}
	\caption{BIND \& \name combined operation: BIND offloads the top $n\in\{0,\cdots, 1\,M\}$ requests from our DNS trace (see Figure~\ref{fig:dnspop}) to \name. As more rules are offloaded, performance increases up to a factor of 5.5.}
	\label{fig:bindmixed}
	\ifthenelse{\equal{\USETIMES}{true}}{\vspace{-1em}}{\vspace{-10pt}}
\end{figure}

Figure~\ref{fig:bindmixed} shows the results as the number of observed replies per second for all four clients for a varying amount of offloaded requests.
When \name is not active (i.e., offloaded entries is 0) we see that BIND performs as in our previous evaluation, serving about 500\,k requests per second.
Once we start offloading the most frequent requests to \name, we observe drastic performance improvements.
Concretely, we see that already offloading the 1\,k most requested hosts increases the overall performance of the server significantly beyond the performance of BIND by 36\,\%.
This speedup factor increases as \name serves more and more requests.
When offloading all available 1\,M entries, \name achieves 2.8\,M replies/s (factor 5.5) in comparison to BIND's 500\,k.
The fact that \name even performs better at 1\,M rules than in the previous setup (1\,M rules, random requests, cf.\, Figure~\ref{fig:eval:alexalinear}) can be attributed to more effective caching of frequently requested rules.

\subsection{Summary}
Based on measured end-user DNS traffic properties, we showed that \name can substantially increase the throughput of a DNS server by a factor of up to 5.5 in mixed scenarios where both BIND and \name are active.
The observed throughput increases can be attributed to reduced per-packet processing times.
Our results thus indicate that the performance of highly loaded DNS servers can be drastically improved by \name.
This cannot only increase the performance on a set of existing hardware, but it can also reduce the hardware requirements for an anticipated workload.
Thus, \name can serve a higher number of requests with less hardware and thereby reduce hardware and energy costs of server deployments.

\section{Evaluation: Memcached}
\label{eval:memcached}
We next move from DNS to a database workload by applying \name to Memcached~\cite{Memcached, fitzpatrick2004distributed} as a prominent example of an in-memory \emph{key-value} (KV) store. 
Such in-memory KV stores are widely used to accelerate the serving of semi-static content, where it is cheaper to cache a value than to re-obtain it.
Website deployments are probably the most popular use-case as reported by Facebook, Amazon, Twitter, or LinkedIn~\cite{atikoglu2012workload, decandia2007dynamo, petrovic2008using, voldemort}, which is why we base our next evaluation on Facebook's Memcached workload patterns.
While the application of Memcached is used as an optimization of current server applications in itself, we now show that we can even further accelerate Memcached by applying \name.

\subsection{Testbed Setup}
For performance evaluation, we use the same testbed as for DNS.
To use Santa, we extended Memcached to install a set of value responses for single keys into \name.
We follow common practice~\cite{nishtala2013} for large installations and use UDP on the transport layer.
Therefore, we adapted DNSperf to send Memcached \texttt{GET} requests and process Memcached responses.
The workload generation is based on a realistic popularity distribution which we discuss next.
While our four clients employ the workload to the server for 45\,s per run, we measure 30\,s after a short warmup for the evaluation.
The test is repeated 30 times.

\subsection{Scenario: Facebook Workload}
To base our evaluation on a realistic scenario, we employ the Facebook workload pattern described by Atikoglu \etal~\cite{atikoglu2012workload}.
The authors present measurement traces and an analysis of Facebook's Memcached deployment.
Overall, there are five different traces from individual caching-pools, each containing keys and values for a separate application or purpose, such as user account information, object metadata or server-side browser information. 
One important property  of all traces is that small keys and values dominate, i.e., most of the keys have a size $\leq$ 32\,B and four out of these five traces mostly contain values $\leq$ 500\,B. 
Moreover, the traces exhibit substantially more \texttt{GET} than \texttt{SET} operations (ratio approximately 30:1).

For our evaluation, we choose the trace containing the user account information. 
This trace contains only two key sizes,~i.e., 16\,B and 21\,B, referencing values with a size of 2\,B.
In this particular trace, the ratio of \texttt{GET} operations in comparison to all operations is 99.8\,\%. 

\begin{figure}[t]
\vspace{-13pt}
{
\subfloat[Cache Hitrate CDF]{\label{fig:cachehit-memcached}\includegraphics[scale=0.92]{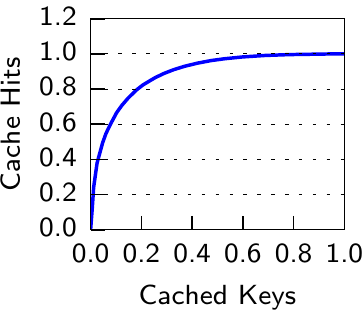}}
\subfloat[Performance results]{\label{fig:perf-memcached}\includegraphics[scale=0.92]{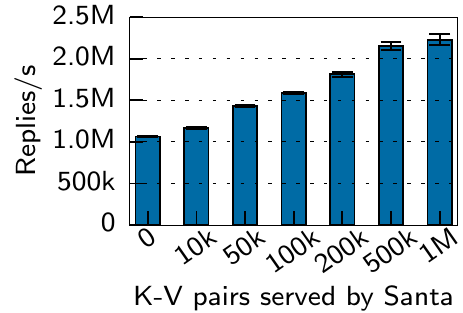}}
\caption{We employ a workload according to the CDF shown in \protect\subref{fig:cachehit-memcached}. This workload is served by Memcached \emph{and} \name in parallel operation: Memcached offloads the top $n\in\{0,\,\ldots,\,1M\}$ ranked requests to \name. As shown in \protect\subref{fig:perf-memcached}, this results in a more significant performance increase the more responses are offloaded to \name.}
\label{fig:memcached-complete}
}
\ifthenelse{\equal{\USETIMES}{true}}{\vspace{-0em}}{\vspace{-1em}}
\end{figure}

Additionally, Atikoglu \etal~\cite{atikoglu2012workload} provide an analysis of the popularity distribution of keys for each individual trace, all showing a power-law distribution.
Based on this information, we derive the achievable cache hit rate shown in Figure~\ref{fig:cachehit-memcached}. 
When offloading just the 5\,\% most popular keys, \name already yields a hit rate of 50\,\%.
Increasing the amount of offloaded keys to 15\,\%, the hit rate increases to 75\,\% due to the power-law distribution.

\subsection{Memcached Performance Evaluation}
We first measure the Memcached baseline performance, i.e., without any \name rule being installed. %
For this and all subsequent measurements, the clients randomly pick keys from a total set with 1\,M keys, weighted by a popularity distribution~\cite{atikoglu2012workload}.
As shown in Figure~\ref{fig:perf-memcached}, Memcached serves 1.06\,M requests per second when $0$ KV pairs are offloaded to \name.

We next show the performance of \name by enabling Memcached to offload KV pairs into our system.
We start by offloading the most frequent 10\,k KV pairs.
Responding to these top 10\,k key requests with \name, the performance increases by a factor of only 1.1.
This is not surprising, as these 10\,k keys only represent 1\,\% of the overall key set in this scenario, leading to a \name-cache hit rate of 22\,\% (cf. Figure~\ref{fig:cachehit-memcached}).
When further increasing the number of \name-answered key requests to 200\,k, i.e., 20\,\% of the overall key set with a \name-cache hit rate of 82\,\%, the performance increases by a factor of 1.7.
Finally, when \name serves all requests (i.e., all existing KV pairs are offloaded to \name), we achieve a rate of 2.2\,M replies per second, resulting in a factor of 2.1.

Besides this UDP evaluation, we have further tested Memcached using long-lived TCP connections (not shown), resulting in a speedup by a factor up to 1.6.
We attribute the lower increase in throughput for TCP as compared to UDP to the heavier TCP state maintenance overhead in the Linux networking stack.

\subsection{Summary}
Based on real-world traces and traffic properties, we have shown that \name can increase the throughput of a Memcached key-value store by a factor of up to 2.1. 
Thus, when having a majority of \texttt{GET} requests on a small subset of keys (e.g., as in the used Facebook workload pattern), \name improves the performance significantly in comparison to a Memcached-only setup.

\section{Evaluation: HTTP Server}
We next employ \name to accelerate HTTP as a prominent, widely-used TCP-based application that currently carries the bulk of traffic in the Internet~\cite{ixpappmix}.

We exemplify this evaluation by focusing on using \name to accelerate the serving of static content, e.g., images, stylesheets, and fonts.
First, these objects are well suited for offloading and second, in an active measurement study we conducted, we observe a range of static content that can generate a substantial amount of traffic.
Optimizing the serving of this content therefore has the potential to reduce general load of servers and empowers the same hardware serving a much larger number of clients.
In this evaluation, we observe that \name can indeed yield large increases in HTTP throughput.
\name therefore provides a promising step in optimizing the overall performance of web applications.

\subsection{Testbed Setup}
We evaluate the TCP performance of \name in the same testbed as used for DNS and Memcached.
Now, we deploy NGINX~\cite{nginx} as a widely-used HTTP server at version 1.9.4 which we extended to install rules into \name. 

We base the workload generation on the popular HTTP benchmarking tool \emph{wrk}~\cite{wrk}.
This tool allows us to control the request pattern, while offering a high performance in terms of requests per second.

\begin{figure}
    \centering
    {\includegraphics[width=\columnwidth]{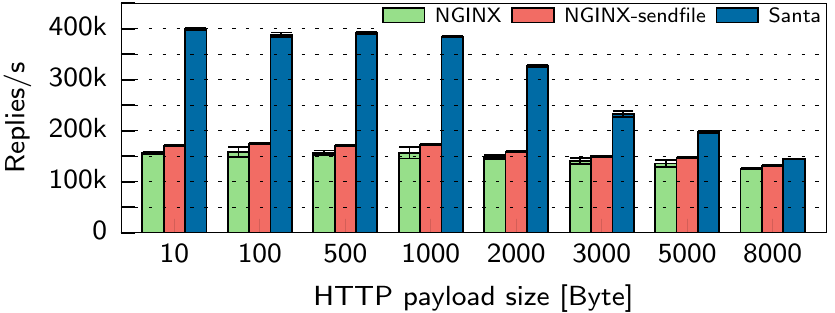}}

    \caption{Comparision of NGINX, NGINX with \texttt{sendfile}, and NGINX with \name. While \texttt{sendfile} slightly improves the NGINX performance, serving replies by \name increases throughput by more than 150\,\% for replies up 1000\,B.}
	\ifthenelse{\equal{\USETIMES}{true}}{\vspace{-0em}}{\vspace{-10pt}}
    \label{fig:tcp:NGINXVsSanta}
\end{figure}

\subsection{Scenario: Frequent Static Web Objects}

We motivate the workload scenario of our HTTP evaluation by measuring frequently accessed static web objects.
Therefore, we crawled the Alexa top 1\,M list with the PhantomJS~\cite{phatomjs} headless browser and analyzed the embedded objects.

Temporarily stable objects include popular APIs, e.g., weather APIs.
Such APIs are frequently accessed from a wide range of devices (e.g., Smartphones) and provide temporarily stable results.
For example, the answer to a weather query only changes once the weather data is updated.
Accelerating the serving of such web objects can increase the overall throughput and thus lower the resource demands of the serving infrastructure.

The ISP trace (see Section \ref{sec:implementation:tcp_processing}) shows that also frequently accessed images, another class of static web objects, are small.
E.g. 80\,\% of the frequently accessed gif images have a size of 43\,B.
Similarly, 55\,\% of all icons and 30\,\% of png graphics are sized less than 1\,kB.

\subsection{HTTP Performance Evaluation}
We start by evaluating the baseline performance of NGINX for different HTTP payload sizes.
To optimize for speed, we configured the NGINX server with commonly applied performance optimizations.
We further distinguish between NGINX using and not using the kernel system call \texttt{sendfile} which allows user space applications to avoid additional copying instructions.
The server provides a single file with a configurable payload size.

We instruct wrk to issue HTTP GET requests for the served file.
After a short warmup period, we measure the performance over 30\,s intervals and repeat each experiment 30~times.
As a performance metric, we measure the served requests per second captured by wrk on the clients.
To assess the combined performance of NGINX and \name, we let NGINX install a \name rule to serve a specific file, thus relieving NGINX by directly processing requests for the specified file from kernel space.

Figure~\ref{fig:tcp:NGINXVsSanta} shows the sum of served requests per second while increasing the HTTP payload size for NGINX, NGINX using \texttt{sendfile}, and \name.
For small payload sizes of up to 1000\,B, the reply rate is stable in all three scenarios.
\texttt{sendfile} increases the NGINX performance by a factor of 1.1, but \name provides better performance by a factor of over \HTTPfactor while being CPU bound in all three scenarios.
When increasing the payload size beyond the typical MSS of 1436\,B, the throughput of \name degrades notably as TCP segments the 2000\,B into two separate packets.
Therefore, a performance drop is expected at each multiple of the MSS as we start spending more and more time in the output path and have to handle additional signaling traffic.
Nevertheless, \name provides significantly higher response rates.
This trend continues until NGINX and \name become equally bounded by the 10\,G line rate at payloads of over 5000\,B.

Even though we offload replies to the kernel, we do not eliminate all system calls.
NGINX still handles \texttt{accept} calls and is notified when the remote-end closes the connection.
Although both system calls can be deferred until \name has matched and replied (\texttt{TCP\_DEFER\_ACCEPT} socket option), they still occur.
It would be possible to close the connection from within the kernel, eliminating all system calls and further increasing the performance, but we refrained from doing so to not tamper with TCP semantics and to not tailor our approach to HTTP.

\afblock{Alternative approaches.}
Placing \name in the kernel stack, as focused on in this work, is only one possible solution.
As illustrated in Figure~\ref{fig:positions} and discussed in Section~\ref{sec:partition}, more specialized solutions exist that bypass the kernel and perform all processing in user space.
One exemplary approach is Seastar~\cite{seastar} offering high performance user space stacks ontop of DPDK~\cite{dpdk} and thus accelerating performance by kernel bypassing.

To get a better understanding of the performance that can be achieved using such a technique, we set up Seastar in our testbed and ran the same HTTP performance tests as for \name and NGINX.
For payloads up to 500\,B, Seastar outperforms vanilla NGINX by a factor of 4. 
Compared to \name, the performance gain is about 1.7.
However, at increased payloads towards link saturation ($\geq$~5000\,B) the difference between \name and Seastar becomes marginal.

Thus, encouraged by the possible performance gain, implementing and evaluating \name on top of approaches like Seastar is a task for our future work.

\subsection{Summary}
We showed that \name can substantially increase the throughput of an HTTP server by a factor of up to \HTTPfactor.
This performance gain is lower than for DNS due to TCP having higher protocol overhead (CPU, bandwidth, and additional context switches).
As in the DNS evaluation, the baseline performance of the NGINX HTTP server is CPU bound and thus cannot fully saturate the 10\,G link for small payload sizes.
While the same is true for \name, it can serve significantly more requests doing so.
For both, \name and NGINX, as long as the CPU is the limiting factor, the amount of replies per second remains steady for smaller payload sizes.
But with increased payload sizes ($\geq$~5000\,B), the 10\,G interface becomes the bottleneck and the performance differences dwindle.
Our results thus indicate that the performance of highly loaded HTTP servers can be drastically improved by \name.

As for DNS and Memcached, the throughput increases can be attributed to reduced per-packet processing times, which also yield latency reductions. 
Depending on the size of the reply, we observe latency improvements at a factor between 4 and 5.
We omit a detailed discussion of these results since typical end-to-end latencies are orders of magnitude higher in Internet deployments.
However, we remark that these latency improvements can become relevant in local settings, e.g., real-time bidding in ad-exchanges that are often upper bounded by 100\,ms~\cite{IMC2014-Back-Office} or in industrial settings, e.g., in-datacenter traffic.

\section{Related Work}
\label{sec:relatedwork}

Classical approaches to optimize packet processing on end-hosts and servers are kernel optimizations and alternative network APIs.
Proposed optimizations involve {\em i)} channelizing processing~\cite{Jacobson06, han2012megapipe}, {\em ii)} alternative socket abstractions~\cite{han2012megapipe}, or {\em iii)} using batching to reduce overheads~\cite{netslices, han2012megapipe}. While these optimizations are application independent, improved packet processing performance can also be obtained by moving the entire server logic into the kernel space.
To this end, classic network processing tasks, steered by the user space, such as firewalling and demultiplexing have long been a kernel feature~\cite{netfilter} often enabled via packet filters~\cite{mogul87} such as BPF~\cite{mccane93}.
Moreover, advancements in NFV make use of this concept of executing user-level code in the kernel environment~\cite{ballani2015,pathak2015}.
Another example application represents the implementation of a kernel-level HTTP cache~\cite{khttpd, lever2000analysis}. 
Such approaches split up static and dynamic HTTP content to be handled by the kernel and a user space web server respectively. %
Serving static content from a kernel space web server can be almost twice as fast as from the user space counterpart~\cite{lever2000analysis}.
While \name benefits from similar performance improvements due to in-kernel packet processing, it is application agnostic and enables every application to offload packet processing tasks expressed via rules. 
To reduce the load on servers based on commonly requested items,~i.e., HTTP or peer-to-peer content,~\cite{sarolahti2011poor} proposes an extension of TCP that allows the content provider to label cacheable items.
Routers on the path cache these labeled packets and serve them directly to clients.
\name is able to process packets based on the aforementioned rules, without the need of protocol modifications for tagging such as labels.
This way, we provide a generalized and application-agnostic framework for offloading packet processing, without altering standardized protocols or requiring on-path assistance.

More radical approaches tackling this challenge involve (partially or completely) bypassing the kernel in the data-plane by either {\em i)} offloading packet processing to specialized hardware, such as GPU based processing~\cite{packetshader, arsenic, vasiliadis2014gaspp, kim2014gpunet} or NetFPGAs~\cite{flajslik2013network}, or by {\em ii)} shifting packet processing to user-land stacks~\cite{netmap, Honda14, sandstorm, jeong2014mtcp}.
The latter represents an active line of research that achieved drastic performance increases and lower CPU footprints by avoiding kernel based packet processing overheads~\cite{sandstorm}.
These advances have proved to be useful for accelerating software switches~\cite{VALE}, HTTP~\cite{Honda14, sandstorm}, and DNS servers~\cite{sandstorm}.
However, while bypassing the kernel largely optimizes the achievable packet processing performance, it usually comes at the cost of abandoning a well-maintained and kernel network stack that offers central administration, although \cite{yasukata16stackmap} reuse the kernel network processing.
Likewise, new OS designs propose to generally remove the kernel from the data-plane~\cite{Arrakis, IX}.

By applying \name to the kernel, we decide to take a different route: Instead of bypassing the kernel stack and performing the packet processing within the application, we enable the offloading of application processing logic into the kernel (or other parts of the network).
This design choice enables backward compatibility for unmodified applications and allows modified applications to benefit from both kernel-accelerated packet processing {\em and} a full-featured kernel stack.
Further, \name is a software solution running on commodity devices, without the need for specialized hardware.

\section{Discussion}
While we discussed the design and implementation decisions and trade-offs in detail in Sections~\ref{sec:architecture} and~\ref{sec:implementation}, some additional general points deserve further discussion.

\afblock{\name vs.\ caching.}
While \name shares similarities with traditional caching, both concepts are fundamentally different.
A traditional cache, be it a CPU, an OS, or a web proxy cache, acts independently of the application.
It manages its cache based on heuristics that try to keep frequently accessed items while removing unpopular or outdated ones.
In contrast, \name's cache is controlled by the application.
It is not fixed to a certain size, and elements are inserted and removed explicitly, not via a cache control algorithm.
Furthermore, updates to outdated information are also triggered by the application itself.
This eliminates disadvantages of caching: First, it prevents outdated information being delivered from the cache.
Second, there is no cache thrashing, due to items not being added and removed based on a popularity metric, and there is need for item eviction in favor of others.

\afblock{Expressiveness of rules.}
\name's rules were deliberately designed to be very simple.
This simplicity allows finding matching rules with high efficiency, as described in Section~\ref{sec:hashing}.
However, relying on a static offset and length for matching on incoming packets does not allow matching on dynamically positioned fields.
One example of such requests are HTTP requests in which one would like to match on specific parts of the request headers, such as sending different replies based on the \texttt{Accept-Encoding} or \texttt{User-Agent}.
Since neither order nor size of the headers are fixed, our current matching implementation fails.
For such a case, a string search algorithm checking for the existence of a match \textit{anywhere} would have to be implemented.
But naturally, more complex matching approaches add more processing overhead.

\afblock{Protocol independence, lower layers.}
Our use-cases show the expected common scenario for employing \name: caching replies for servers of well-known application-layer protocols.
However, since \name rules are protocol agnostic and can match on arbitrary parts of incoming packets, there is no need to speak any already existing protocol.
In fact, \name also supports hooking rules earlier into the packet processing than presented in this paper (ontop of UDP/TCP), for example, on the network layer.
But matching ealier in the protocol inhibits a distinction between application sockets, providing the possibility to hijack packets and therefore, should require root access, similar to opening raw sockets.

\afblock{Encrypted Traffic.}
A continuously increasing amount of Internet traffic is protected against eavesdropping and alteration~\cite{Naylor2014cost}---e.g., as expressed by the growth in the HTTPS ecosystem~\cite{httpsIMC}---and can challenge packet matching performed in kernel space.
Concretely, encryption that is applied {\em below} the transport layer, e.g., IPsec or link-layer encryption, does not affect \name since the decryption happens before the matching.
However, encryption that is applied {\em above} the transport layer (e.g., TLS~\cite{rfc5246}) poses a particular challenge to \name.
Since decryption is performed in user space, packet matching cannot be done in kernel space.
Although methods for matching on encrypted traffic (e.g.,~\cite{sherry2015blindbox}) or enabling a selected party to access and modify traffic (e.g.,~\cite{naylor2015mctls}) exist, these add considerable overhead and require modification of clients.
To address this challenge, we plan to integrate TLS en- and decryption on the transport layer in collaboration with cryptographic libraries,~e.g. OpenSSL~\cite{OpenSSL}, to allow \name to match on encrypted traffic as part of future work.

\afblock{Enabling in-network processing.}
While our approach focuses on partially offloading application logic to the kernel space for faster processing, it is not limited to in-kernel processing.
Concretely, application logic expressed by simple SDN-inspired \name rules can be conceptually executed in {\em any} network device, e.g., ranging from the NIC to edge switches.
Thus, this paves the way for enabling lightweight in-network processing approaches.

\section{Conclusion}
This paper describes \name, a new approach to accelerate server systems by proposing a widely applicable data path architecture that can benefit from current bypassing approaches, but does not necessarily abandon well-maintained kernel stacks.
\name allows applications to offload their frequent packet processing tasks into an application-agnostic rule-processor that can reside in various parts of the network, e.g., in an OS kernel.
It thus essentially shortens the data path.
We demonstrate \name's potential by implementing it in the Linux network stack.
Our evaluation of a broad class of UDP and TCP based applications (DNS, Memcached, and HTTP), each motivated by real-world traffic pattern, highlights that \name can increase the number of served requests by a factor of \MEMfactor to \DNSfactor, depending on the application.

By the example application of \name to the kernel, we show that significant performance increases can be reached.
Furthermore, while we focused on this domain, \name's design is not limited to the kernel and thus presents a first step for enabling lightweight in-network processing.
This protocol-agnostic approach, despite its ease of use, opens up the possibility for performance improvement to all applications that struggle under heavy load from static request--response pairs.
Thus, it can both increase potential throughput on existing machines and reduce the number of servers required to handle the same workload.

\bibliographystyle{abbrv}
\bibliography{references} 

\begin{thebibliography}{10}
\bibitem{santa-source}
{Santa source code will be published upon acceptance of the paper}.
\end{thebibliography}

\balance

\end{document}